\documentclass[runningheads]{llncs}
\usepackage{times}
\usepackage{graphicx}
\usepackage{latexsym}
\usepackage{helvet}
\usepackage{courier}
\usepackage{amsmath}
\usepackage{amssymb}
\usepackage{algorithm}
\usepackage{alltt}
\usepackage{verbatim}
\usepackage{url}
\usepackage{xspace}
\usepackage[defblank]{paralist}

\usepackage{pgfplots}
\pgfplotsset{compat=1.14}
\usepackage{tikz}
\usepackage[caption=false]{subfig}

\usepackage{algorithmicx}
\usepackage{algpseudocode}

\newcommand{\ignore}[1]{}
\newcommand{\finish}[1]{}
\newcommand{\skipit}[1]{{ #1}}

\newcommand{\floor}[1]{\lfloor #1 \rfloor}

\newcommand{\DL}{{ DL}}

\newcommand{\pl}{\partial_{||}}

\newcommand{\PD}[1]{+\Delta #1}

\newcommand{\ARROW}{\hookrightarrow}

\newcommand{\non}{{\thicksim}}

\renewcommand{\imath}{i}

\newcounter{clause}
\def\theclause{$c$\arabic{clause}}

{\end{tabbing}}

{\end{tabbing}}

\newcommand{\mi}[1]{\mathit{#1}}

\begin{document}

\title{Which are the True Defeasible Logics?}

\author{Michael J. Maher}   
\institute{
Reasoning Research Institute \\
Canberra, Australia  \\
E-mail: michael.maher@reasoning.org.au
}
\authorrunning{M.J. Maher}

%

\maketitle

\begin{abstract}
The class of defeasible logics is only vaguely defined --
it is defined by a few exemplars and the general idea of efficient reasoning with defeasible rules.
The recent definition of the defeasible logic $\DL(\pl)$ introduced new features to such logics,
which have repercussions that we explore.
In particular, we define a class of logics that accommodates the new logic while retaining the 
traditional properties of defeasible logics.

\end{abstract}

\section{Introduction}

Defeasible logics were introduced by Nute \cite{Nute88,NuteINAP}
as logics in which to reason about rules of thumb,
as well as statements qualified by terms such as
``generally'', ``typically'', ``in the main'' and so on.
Such statements may provide \emph{prima facie} evidence for a conclusion 
based on specific facts but,
by their very nature, may be subject to exceptions.
Thus the rules are \emph{defeasible}:
unsound, in general, even if usually valid.

Despite the unsoundness, such rules of thumb are ubiquitous.
They occur in everyday speech,
even if the archetypal rule of thumb  ``birds fly'' does not arise every day.
Except for definitions, very few, if any, statements hold without exception.
Defeasible rules are also used to represent legal documents and legal reasoning
\cite{Prakken,regs,contracts,Grosof04}\footnote{
Indeed, 
legal rules are generally considered defeasible,
although there is debate on whether legal rules are necessarily defeasible \cite{Hart} or not \cite{Schauer}.
} and in business process compliance \cite{BPC}.
They have also been suggested as an ideal basis for representing and explaining
conclusions drawn by machine learning systems \cite{CT16,StrassWD19,VBP21,boardgameqa}.

The inference rules of a defeasible logic provide ways in which
these rules of thumb can be applied to known facts to infer statements that may be considered
justified.
That is, inference rules  provide a formalisation of how the defeasibility of the rules
is applied in the process of drawing conclusions.
They can provide ways to deal with conflicting rules of thumb,
preventing the inference of contradictory statements.
To achieve appropriate resolution of conflicts, inference rules in defeasible logics
are substantially more complicated than inference rules in most logics.
This provides a wide range of potential inference rules for defeasible logics.

Several concrete defeasible logics have been studied in detail
(see, for example, \cite{TOCL10,MN10,BillingtonPPL} and references therein),
as well as general frameworks \cite{Nute_book,flexf}.
Nevertheless, there has been no investigation of what counts as a defeasible logic.
This was thrown into sharp relief by the logic defined in \cite{sdl} which included novel features
in the definition of its inference rules.
The undisciplined use of such features can lead to logics that veer away from the
original intent of defeasible logics:
to draw conclusions from facts and defeasible rules in a principled and computationally efficient way.
Consequently, it becomes necessary to identify restrictions on the use of these features that
will preserve the traditional capabilities of defeasible logic.

In this paper we define a class of well-behaved defeasible logics
that satisfies this requirement.
This class is motivated by
existing desiderata for defeasible logics,
such as coherence and a principle of strong negation,
as well as a desire to retain a simple proof theory.
It accommodates many existing defeasible logics.
A signal advantage of our approach is that it is largely agnostic to the syntax of defeasible theories.

The culmination of the paper is a proof that every well-behaved defeasible logic is coherent.
Coherence is a vital property for defeasible logics:
without it, the conventional interpretation of negative tags is unsustainable.
Although variants of this theorem have been claimed before,
the class of defeasible logics to which it applied was unclear.
This provided one of the motivations for this paper.

The remainder of the paper is structured as follows.
Section \ref{sect:defeasible_logic} introduces
the necessary preliminaries on defeasible logics.
It includes definitions of the logics $\DL(\partial)$  \cite{TOCL01} and $\DL(\pl)$ \cite{sdl},
but readers familiar with these logics can skip this part.
It also includes the definition of some significant properties of defeasible logics. 
The next section identifies some problems with logics that
naively incorporate elements of $\DL(\partial)$ and $\DL(\pl)$ in their inference rules
-- problems that make these logics unsuitable for the original purpose of defeasible logics.
In response, Section~\ref{sect:WDL}
explores restrictions on the inference rules used in logics to preserve desirable properties
and avoid the problems.
Building on this work, Section~\ref{sect:WBDL} proposes
a well-behaviour criterion for "true" defeasible logics,
and shows that all well-behaved defeasible logics are coherent.
Well-behaviour is based on properties of inference rules and proofs
and their relation to desirable properties of defeasible logics.
This section also provides a syntactically-defined subset of well-behaved defeasible logics.
The paper closes in Section~\ref{sect:conc}
with a summary of the achievements of this paper
and a brief indication of topics that deserve further investigation.

\section{Defeasible Logics}  \label{sect:defeasible_logic}

This section can be skipped by readers familiar with defeasible logics.
The only variation from other presentations of defeasible logics is that
inference rules are presented as such, rather than as conditions on proofs.

\subsection{Defeasible Theories}  \label{sect:DT}

A defeasible theory $D$ is a triple $(F,R,>)$ where $F$ is a finite set of facts (literals), 
$R$ a finite set of labelled rules,
and $>$ a superiority (or priority) relation (a binary acyclic relation) on $R$ (expressed on the labels),
specifying when one rule overrides another, given that both are applicable.

A rule $r$ consists (a) of its antecedent (or body) $A(r)$ which is a finite set of literals, (b) an arrow, and, (c) its
consequent (or head) $C(r)$ which is a literal. Rules also have distinct \emph{labels} which are used to refer to
the rule in the superiority relation. There are three types of rules: strict rules, defeasible rules and
defeaters represented by a respective arrow $\rightarrow$, $\Rightarrow$ and $\leadsto$. Strict rules are rules in
the classical sense: whenever the premises are indisputable (e.g., facts) then so is the conclusion. Defeasible rules
are rules that can be defeated by contrary evidence. Defeaters are rules that cannot be used to draw any conclusions; their only use is to provide contrary evidence that may prevent some conclusions.
We use $\ARROW$ to range over the different kinds of arrows used in a defeasible theory.

We will address only propositional logics in this paper.
Nevertheless, some examples will be written with a first-order syntax;
each rule represents the set of its variable-free instances, and each variable-free atom is considered a proposition.
A literal is either a proposition or its negation.
Negation is represented by the symbol $\neg$,
and we define the complement operation $\non$ as follows:
if $q$ is a proposition $p$, then $\non q = \neg p$;
if $q$ has the form $\neg p$ then $\non q = p$.
Given a set $R$ of rules, we denote the set of all strict rules in $R$ by
$R_{s}$, and the set of strict and defeasible rules in $R$ by $R_{sd}$. 
$R[q]$ denotes the set of rules in $R$ with consequent $q$.

\begin{example}    \label{ex:tweety}
To demonstrate defeasible theories,
we consider the familiar Tweety problem and its representation as a defeasible theory.
The defeasible theory $D$ consists of the rules and facts
\[
\begin{array}{rrcl}
r_{1}: & \mi{bird}(X) & \Rightarrow & \phantom{\neg} \mi{fly}(X) \\
r_{2}: & \mi{penguin}(X) & \Rightarrow & \neg \mi{fly}(X) \\
r_{3}: & \mi{penguin}(X) & \rightarrow & \phantom{\neg} \mi{bird}(X) \\
r_{4}: & \mi{injured}(X)   & \leadsto     &  \neg \mi{fly}(X) \\
f      : & \mi{penguin}(\mi{tweety}) &  &  \\
g      : & \mi{bird}(\mi{freddie}) &  &  \\
h     : & \mi{injured}(\mi{freddie}) &  &  \\
\end{array}
\]
and a priority relation $r_{2} > r_{1}$.

Here $r_1, r_2, r_3, r_4, f$ are labels and
$r_3$ is (a reference to) a strict rule, while $r_1$ and $r_2$ are defeasible rules,
$r_4$ is a defeater,
and $f, g, h$ are facts.
Thus $F = \{f,g,h\}$, $R_s = \{ r_3 \}$, $R_{sd} = R = \{r_1, r_2, r_3 \}$,
and $>$ consists of the single tuple $(r_2, r_1)$.
The rules express that birds usually fly ($r_1$),
penguins usually don’t fly ($r_2$),
that all penguins are birds ($r_3$),
and that an injured animal may not be able to fly ($r_4$).
In addition, the priority of $r_{2}$ over $r_{1}$ expresses that when something is both a bird and a penguin (that is, when both rules are ready to fire) it usually cannot fly
(that is, only $r_{2}$ may fire, it overrules $r_{1}$).
Finally, we are given the facts that $\mi{tweety}$ is a penguin,
and $\mi{freddie}$ is an injured bird.
\end{example}

A \emph{defeasible logic} defines how conclusions may be drawn from a defeasible theory.
A \emph{conclusion} takes the forms $+d \, q$ or $-d \, q$, where $q$ is a literal and $d$ is a tag indicating which inference rule was used.
Intuitively, given a defeasible theory $D$,
$+d \, q$ expresses that $q$ can be proved via inference rule $d$ from $D$,
while $-d \, q$ expresses that it can be established that $q$ cannot be proved from $D$ via $d$.
Formally, we write $D \vdash +d \, q$ to express that $+d \, q$ can be proved from $D$
(and similarly for $-d \, q$).
In this case we say that $+d q$ ($-d q$) is a \emph{consequence}
of $D$ in the defeasible logic.
Thus a consequence is a valid conclusion in the application of the logic to the defeasible theory $D$.

Proofs are sequences of consequences where each consequence is derived by an inference rule
from the preceding consequences in the proof.
We will use $\in$ for both membership of a sequence and membership of a set.
Inference rules for defeasible logics are usually expressed as conditions on proofs,
but in this paper we will present them as the ability to append a conclusion to a proof\footnote{
The reasons for this change are to give readers a more familiar format for the logics
and to simplify the syntax for inference rules.
}.
For the two logics that follow, that presentation varies from their original presentation
but this is only a surface change.
In general, an inference rule has the form
\begin{center}
We may append  $\pm d \, q$ to $P$ if $C$
\end{center}
where $\pm$ may be $+$ or $-$, $q$ is a literal, $P$ is a proof, and $C$ is an \emph{applicabiity condition},
a Boolean-valued formula restricting when the inference may be drawn.
$C$ is parameterized by $\pm d$, $q$, $P$, and $D$, the defeasible theory to which the inference rule is applied.
Parameters of $C$ will be omitted when not relevant.

Defeasible logics are usually named $\DL(d)$, where $d$ is the tag of the main inference rule.

\subsection{The Logic $\DL(\partial)$}

For example, in \cite{TOCL01}, a defeasible logic, now called $\DL(\partial)$, is defined
with the following inference rules\footnote{
Here, 
$D$ is a defeasible theory  $(F,R,>)$,
$q$ is a literal, and
$P$ denotes a proof.
As noted earlier,
$X[q]$ denotes the subset of a set of rules $X$ with head $q$,
$R_s$ denotes the set of strict rules in $R$,
and $R_{sd}$ denotes the set of strict or defeasible rules (i.e. non-defeaters) in $R$.
}

\noindent\begin{minipage}[t]{.45\textwidth}
\begin{tabbing}
90123456\=7890\=1234\=5678\=9012\=3456\=\kill

$+\Delta)$  We may append  $+\Delta q$ to $P$ if \\
\hspace{0.2in}  (1)  $q \in F$;  or \\
\hspace{0.2in}  (2)  $\exists r \in R_{s}[q] \  \forall a \in A(r),
+\Delta a \in P$.
\end{tabbing}
\end{minipage}
\begin{minipage}[t]{.45\textwidth}
\begin{tabbing}
90123456\=7890\=1234\=5678\=9012\=3456\=\kill

$-\Delta)$  We may append  $-\Delta q$ to $P$ if \\
\hspace{0.2in}  (1)  $q \notin F$,  and \\
\hspace{0.2in}  (2)  $\forall r \in R_{s}[q] \  \exists a \in A(r),
-\Delta a \in P$.
\end{tabbing}
\end{minipage}
\smallskip
\smallskip

These two inference rules concern reasoning about definitive information,
involving only strict rules and facts.
They define conventional monotonic inference ($+\Delta$)
and provable inability to prove from strict rules and facts ($-\Delta$).
Most defeasible logics contain these inference rules, along with others;
we will assume, throughout the paper, that all logics discussed contain these inference rules.
The logic involving just these inference rules is denoted by $\DL(\Delta)$.
The next rules refer to defeasible reasoning.

\smallskip
\smallskip
\noindent\begin{minipage}[t]{.45\textwidth}
\begin{tabbing}
$+\partial)$  We may append  $+\partial q$ to $P$ if \\
\hspace{0.2in}  (1)  $+\Delta q \in P$; or  \\
\hspace{0.2in}  (2)  The following three conditions hold. \\
\hspace{0.4in}      (2.1)  $\exists r \in R_{sd}[q] \  \forall a \in A(r)$,  \\
\hspace{1.1in}                                  $+\partial a \in P$,  and \\
\hspace{0.4in}      (2.2)  $-\Delta \non q \in P$,  and \\
\hspace{0.4in}      (2.3)  $\forall s \in R[\non q]$  either \\
\hspace{0.6in}         (2.3.1)  $\exists a \in A(s),  -\partial a \in P$;  or \\
\hspace{0.6in}          (2.3.2)  $\exists t \in R_{sd}[q]$  such that \\
\hspace{0.8in}                $\forall a \in A(t),  +\partial a \in P$,  and \\
\hspace{0.8in}                $t > s$.
\end{tabbing}
\end{minipage}
\begin{minipage}[t]{.45\textwidth}
\begin{tabbing}
$-\partial)$  We may append  $-\partial q$ to $P$ if \\
\hspace{0.2in}  (1)  $-\Delta q \in P$, and \\
\hspace{0.2in}  (2)  either \\
\hspace{0.4in}      (2.1)  $\forall r \in R_{sd}[q] \  \exists a \in A(r)$,  \\
\hspace{1.1in}                                        $-\partial a \in P$; or \\
\hspace{0.4in}      (2.2)  $+\Delta \non q \in P$; or \\
\hspace{0.4in}      (2.3)  $\exists s \in R[\non q]$  such that \\
\hspace{0.6in}          (2.3.1)  $\forall a \in A(s),  +\partial a \in P$,  and \\
\hspace{0.6in}          (2.3.2)  $\forall t \in R_{sd}[q]$  either \\
\hspace{0.8in}                $\exists a \in A(t),  -\partial a \in P$;  or \\
\hspace{0.8in}                not$(t > s)$.\\
\end{tabbing}
\end{minipage}

In the $+\partial$ inference rule,
(1) ensures that any monotonic consequence is also a defeasible consequence.
(2) allows the application of a rule (2.1) with head $q$, provided that
monotonic inference provably cannot prove $\non q$ (2.2)
and every competing rule either provably fails to apply (2.3.1)
or is overridden by an applicable rule for $q$ (2.3.2).

In the $-\partial$ inference rule,
$q$ must be (1) monotonically provably unprovable, and either
(2.1) the bodies of all rules for $q$ are provably unprovable,
(2.2) $\non q$ is proved monotonically, or
(2.3) there is a competing rule $s$ whose body is proved
and every rule for $q$ either has provably unprovable body
or cannot override $s$.

\begin{example}    \label{ex:tweety2}
The above inference rules make several inferences
from the Tweety defeasible theory in Example \ref{ex:tweety}. 

The $+\Delta$ inference rule infers
$+\Delta\, \mi{penguin}(\mi{tweety})$,
$+\Delta\, \mi{bird}(\mi{freddie})$, and
$+\Delta\, \mi{injured}(\mi{freddie})$ 
from the facts, and
$+\Delta\, \mi{bird}(\mi{tweety})$
using $r_3$.
Such inferences are definite conclusions from the theory.
The $-\Delta$ inference rule infers, among others
$-\Delta\,  \mi{penguin}(\mi{freddie})$,
$-\Delta\, \mi{injured}(\mi{tweety})$,
$-\Delta\, \neg \mi{injured}(\mi{tweety})$, and
$-\Delta\, \neg \mi{bird}(\mi{tweety})$,
indicating that the theory is provably unable to come to a definite conclusion about these statements,
because there is no rule (and no fact) for these literals.
It also infers
$-\Delta\,  \mi{fly}(\mi{freddie})$,
$-\Delta\,  \neg \mi{fly}(\mi{freddie})$,
$-\Delta\,  \mi{fly}(\mi{tweety})$, and
$-\Delta\, \neg  \mi{fly}(\mi{tweety})$
because there is no strict rule for $\mi{fly}$ or $\neg \mi{fly}$
and consequently (2) of the $-\Delta$ inference rule is vacuously true.

The $+\partial$ inference rule infers
$+\partial\, \mi{penguin}(\mi{tweety})$,
$+\partial\, \mi{bird}(\mi{freddie})$, and
$+\partial\, \mi{injured}(\mi{freddie})$, and 
$+\partial\, \mi{bird}(\mi{tweety})$
because these statements are known definitely.
It also concludes 
$+\partial\,  \neg  \mi{fly}(\mi{tweety})$
using rule $r_2$ in (2.1),
the previous conclusion 
$-\Delta\,  \mi{fly}(\mi{tweety})$ in (2.2),
and, despite the presence of $r_1$ as $s$ in (2.3),
using $r_2$ as $t$ in (2.3.2) with the priority statement $r_{2} > r_{1}$
to overrule $r_1$.
It is unable to similarly conclude 
$+\partial\,  \mi{fly}(\mi{freddie})$,
because of the presence of $r_3$ and the lack of a priority statement to overrule it.

The $-\partial$ inference rule infers, among others,
$-\partial\,  \mi{penguin}(\mi{freddie})$ and
$-\partial\, \mi{injured}(\mi{tweety})$
because these statements are known unprovable definitely (1),
and (2.1) is satisfied vacuously because there is no rule for these predicates.
It also infers
$-\partial\,  \mi{fly}(\mi{freddie})$,
$-\partial\,  \neg \mi{fly}(\mi{freddie})$, and
$-\partial\,  \mi{fly}(\mi{tweety})$.
\end{example}

\vspace{0.2cm}

\subsection{The Logic $\DL(\pl)$}\label{sect:sdl}

The defeasible logic  $\DL(\pl)$  \cite{sdl} was designed to allow defeasible inference
to be scalable to very large data sets.
We present the inference rules of that logic here.

$\DL(\pl)$ involves three tags: 
$\Delta$, which expresses conventional monotonic inference;
$\lambda$, an auxiliary tag;
and $\pl$, which is the main notion of defeasible proof in this logic.
The inference rules are presented below\footnote{
As in the inference rules for $\DL(\partial)$ in the previous section,
$D$ is a defeasible theory  $(F,R,>)$,
$q$ is a variable-free literal, and
$P$ denotes a proof.
}.

The monotonic inference rule $+\Delta$, as seen before in $\DL(\partial)$, 
also appears in  $\DL(\pl)$.
For a defeasible theory $D$,
we define $P_{+\Delta}$ to be 
the set of consequences in a largest proof without duplications,
satisfying the applicability condition $+\Delta$ at each step,
and call this the $+\Delta$ \emph{closure}\footnote{
We depart slightly from \cite{sdl} in the notation for closures,
to fit with later notation.
}.
It contains all $+\Delta$ consequences of $D$.

Once $P_{+\Delta}$ is computed, we can apply the $+\lambda$ inference rule.
$+\lambda q$ is intended to mean that $q$ is potentially defeasibly provable in $D$.
The $+\lambda$ inference rule is as follows.

\begin{tabbing}
$+\lambda$: \= We may append  $+\lambda q$ to $P$ if  \\
\> (1) \=$+\Delta q \in P_{\Delta}$ or \\
\> (2)	\>(2.1) $\exists r \in R_{sd}[q] ~ \forall \alpha \in A(r): +\lambda \alpha \in P$ and \\
\> \>(2.2) $+\Delta \non q \notin P_{\Delta}$ 
\end{tabbing}
\smallskip

Using this inference rule, and given $P_{+\Delta}$, we can compute the $\lambda$ closure $P_{+\lambda}$.
which contains all $+\lambda$ consequences of $D$.

$+\pl q$ is intended to mean that $q$ is defeasibly provable in $D$.
Once $P_{+\Delta}$ and $P_{+\lambda}$ are computed, we can apply the $+\pl$ inference rule.

\begin{tabbing}
$+\pl$: \= We may append  $+\pl q$ to $P$ if \\
\> (1) \=$+\Delta q  \in P_{+\Delta}$ or \\
\> (2)	\>(2.1) $\exists r \in R_{sd}[q] ~ \forall \alpha \in A(r): +\pl \alpha \in P$ and \\
\> \>(2.2) $+\Delta \non q \notin P_{+\Delta}$ and \\
\> \>(2.3) \=$\forall s \in R[\non q]$ either \\
\> \> \>(2.3.1) $\exists \alpha \in A(s): +\lambda \alpha \notin P_{+\lambda}$ or \\
\> \>\>(2.3.\=2) $\exists t \in R_{sd}[q]$ such that \\
\> \>\>\>$\forall \alpha \in A(t): +\pl \alpha \in P$ and $t > s$
\end{tabbing}

The $\pl$ closure $P_{+\pl}$ contains all $+\pl$ consequences of $D$.
We say a set of tagged literals $Q$ is \emph{$+\pl$-deductively closed} if,
given closures $P_{+\Delta}$ and $P_{+\lambda}$ as defined above,
for every literal $q$ that
may be appended to $Q$ by the inference rule $+\pl$, $q \in Q$.
Clearly, $P_{+\pl}$ is the smallest $+\pl$-deductively closed set.

\begin{example}    \label{ex:tweety3}
We now apply these
inference rules to the Tweety defeasible theory in Example \ref{ex:tweety}. 

As before, the $+\Delta$ inference rule infers
$+\Delta\, \mi{penguin}(\mi{tweety})$,
$+\Delta\, \mi{bird}(\mi{freddie})$, and
$+\Delta\, \mi{injured}(\mi{freddie})$ 
from the facts, and
$+\Delta\, \mi{bird}(\mi{tweety})$
using $r_3$.
We have no need of the $-\Delta$ inference rule.

Using the $+\lambda$ inference rule
we infer all the literals inferred by the $+\Delta$ inference rule as $+\lambda$ conclusions.
In addition, the rule infers
$+\lambda\, \mi{fly}(\mi{tweety})$,
$+\lambda\, \mi{fly}(\mi{freddie})$, and
$+\lambda\, \neg\mi{fly}(\mi{tweety})$.

Using the $+\pl$ inference rule,
again all the $+\Delta$ conclusions are inferred as $+\pl$ conclusions.
The only other conclusion that can be drawn with this rule is
$+\pl\, \neg\mi{fly}(\mi{tweety})$.
The potential inference of $\mi{fly}(\mi{tweety})$ is overruled by $r_2$ inferring $+\pl\, \neg\mi{fly}(\mi{tweety})$.
On the other hand, a potential inference of $\mi{fly}(\mi{freddie})$ is not obtained
because $r_2$ cannot overrule $r_4$.
\end{example}

A key feature of $\DL(\pl)$ 
inference rules is that
they do not use negative inference rules, unlike  $\DL(\partial)$ which uses $-\Delta$ and $-\partial$.
Instead they use expressions $+\Delta q \notin P_{+\Delta}$ and $+\lambda q \notin P_{+\lambda}$.
This choice was based on practical difficulties in 
scalably implementing existing non-propositional defeasible logics \cite{sdl,ECAI2012}.
It provides structural benefits when these logics are translated into logic programs \cite{sdl2}.
Technical relationships between $\DL(\partial)$ and $\DL(\pl)$ are established in \cite{sdl,sdl3}.
Although negative inference rules are unnecessary in $\DL(\pl)$, 
they are available in Appendix C of \cite{sdl}.

\subsection{Properties of Defeasible Logics}

Defeasible logics are, in general, non-monotonic:
a defeasible theory $D$ might have a consequence $+d \; q$,
but the addition of information to $D$ can result in this no longer being a consequence.
In terms  of non-monotonic reasoning, traditionally defeasible logics have employed an approach 
referred to as ``directly skeptical'' \cite{HTT90,Horty02}.
As a result, the consequences of a defeasible theory in a given logic form
a single extension (in contrast to default logic \cite{DefaultLogic}).
This is a part of the reason why propositional defeasible logics generally
have polynomial cost \cite{linear,TOCL10} to compute the consequences,
while other non-monotonic logics are higher in the polynomial hierarchy.

However, the most important property of a defeasible logic is coherence.
The next definitions are modified slightly from \cite{TOCL10} to account for logics with auxiliary tags,
such as $\lambda$ in $\DL(\pl)$, and logics with multiple forms of defeasible reasoning, 
as in annotated defeasible logic \cite{adl}.

A set $S$ is \emph{coherent} if, for no tag $d$ and literal $q$ does $\{+d q, -d q\} \subseteq S$.
A tag $d$ is \emph{coherent} if, for every defeasible theory $D$  and
every literal $q$, we do not have both $D \vdash +dq$ and $D \vdash -dq$.
A logic is \emph{coherent} if its main inference rules are coherent.
Coherence is essential for defeasible logics,
since no literal should be both provable and (provably) unprovable at the same time.
$\DL(\partial)$ and $\DL(\pl)$ are coherent.

A tag/inference rule $+d$ in a logic containing the $+\Delta$ inference rule is \emph{consistent} if,
for every defeasible theory $D$ in the logic
and every proposition $q$,
we do not have both consequences
$+d q$
and
$+d\neg q$
unless we also have consequences
$\PD{q}$
and
$\PD{\neg q}$.
This property expresses that defeasible reasoning does not cause inconsistencies:
any inconsistency in consequences is caused by inconsistency in the monotonic part of the defeasible theory. 
Thus it is a kind of paraconsistency.
We say a logic is consistent if its main inference rules are consistent.
$\DL(\partial)$ and $\DL(\pl)$ are consistent (even though $\Delta$ and $\lambda$ are not consistent), 
as are the other logics in \cite{TOCL10}.
Notice that consistency does not consider consequences of the form $-d q$ because
having both $-d q$ and $-d \non q$ 
is a matter of having too little information, rather than too much,
and is not unusual in defeasible reasoning.

The closures in the previous subsection are simple because only one inference rule is involved each time.
In general, there may be multiple inference rules defined mutually recursively.
So we now formulate closure in this more general setting.

There are two approaches to defining a closure:
one is based on the smallest closed set of conclusions,
while the other is built on the longest proof without duplicates.

A set $I$ of inference rules is \emph{reference-closed} if, for every $i \in I$,
if the applicability condition $C_i$ for $i$ refers to membership of a conclusion $c$ with tag $i'$ in the proof $P$,
then $I$ also contains $i'$.
For each defeasible logic and set $J$ of inference rules there is a unique smallest reference-closed set
containing $J$.

Given a reference-closed set $I$ of inference rules, a set $S$ of conclusions is \emph{$I$-closed} if
every conclusion that can be inferred by an inference rule in $I$ from a sequence of elements of $S$,
is already in $S$.
The smallest $I$-closed set is called the \emph{$I$-inferential closure}.

In general, not all conclusions in an $I$-closure are of interest:
many may derive from auxiliary inference rules that are needed in the chain of inferences
but are not of interest in themselves.
For example,  $\DL(\partial)$ involves four inference rules, 
but we might only be interested in $+\partial$ conclusions.
Hence we need to define closure for a subset of $I$.

\begin{definition}   \label{defn:closure}
Let $J$ be a set of inference rules and  $I$ be the smallest reference-closed set containing $J$.
Then we define the \emph{$J$-closure} to be
$\{  j \, q ~|~   j \in J,    j \, q  \in X _I \}$, where $X_I$ is the $I$-inferential closure.
\end{definition}

Alternatively, 
let $T$ be the largest proof using only inference rules from $I$ that avoids any repetition of conclusions\footnote{
If repetitions are allowed there is no limit on the length of proofs.
}.
We can then define the $J$-closure as $\{  j \, q ~|~   j \in J,    j \, q  \in T \}$.
If all inference rules in $I$ are monotonic then $T$, considered as a set, is equal to the $I$-inferential closure.
Thus, in that case, the two definitions of $J$-closure are equivalent.
An inference rule is \emph{monotonic} if every conclusion it can infer from a set $U$
can also be inferred from any set $V$ with $U \subseteq V$.

To simplify notation,
we will write $P_d$ for the $\{+d, -d\}$-closure
(so that, for example, $P_\Delta$ refers to the  $\{+\Delta, -\Delta \}$-closure).

Although these definitions are suitable for $\DL(\pl)$, $\DL(\partial)$, and the other logics in \cite{TOCL10}
(all inference rules in these logics are monotonic),
they can be ill-defined for a larger class of logics.
Among the issues:
the mixing of sets and proofs (which are sequences), 
repeated conclusions are sometimes necessary in proofs,
the largest proof might not contain all consequences in a logic,
and there might not be a unique smallest $I$-closed set.
These issues will need to be resolved if closures are to be used as part of
a larger class of defeasible logics.

\subsection{The Principle of Strong Negation}  \label{sect:PSN}

The Principle of Strong Negation (PoSN) was introduced in \cite{flexf}
to recognise a pattern in defeasible logics in which the inference rules for $+d$ and $-d$
have a similar structure.
The applicability condition of the $-d$ inference rule can be seen as a kind of negation of the applicability condition of the $+d$ inference rule.
It can be thought of as a systematic identification of the ways in which the $+d$ inference rule
can fail to apply.
The inference rules for $+\Delta$ and $-\Delta$,
and $+\partial$ and $-\partial$, are examples of this pattern.
Not all defeasible logics have this form,
and the framework of \cite{Nute_book} admits logics that violate this principle.
Nevertheless, most defeasible logics essentially satisfy this principle.

The \emph{strong negation} of an applicability condition was expressed in \cite{flexf} by the function $sneg$
defined inductively as follows:
\[
\begin{array}{l l l}
sneg(+d{p} \in X)  & = &  -d{p} \in X \\
sneg(-d{p} \in X)  & = &  +d{p} \in X \\
sneg( A \wedge B )  & = &  sneg( A ) \vee sneg( B ) \\
sneg( A \vee B )  & = &  sneg( A ) \wedge sneg( B ) \\
sneg( \exists x ~ A )  & = &  \forall x ~ sneg( A )  \\
sneg( \forall x ~ A )  & = &  \exists x ~ sneg( A )  \\
sneg( \neg A )  & = &  \neg sneg( A )  \\
sneg( A )  & = &  \neg A \hspace{1.1in} \mbox{if $A$ is a pure formula}\\
\end{array}
\]
A \emph{pure formula} is a formula that does not contain a tagged literal.
All tags $d$ are treated as in the first two equations.
The strong negation of the
applicability condition of an inference rule is a constructive
approximation of the conditions where the rule is not applicable.

The \emph{Principle of Strong Negation} requires that,
if we have an inference rule
\begin{center}
We may append  $+d \, q$ to $P$ if $C$
\end{center}
then the inference rule for $-d$ should be
\begin{center}
We may append  $-d \, q$ to $P$ if $sneg(C)$
\end{center}
We say that a logic \emph{supports} the Principle if
every tag $d$ in the logic satisfies the Principle.

An advantage of the definition of $sneg$ is that it is a self-inverse,
that is, for any condition $X$, $sneg( sneg( X ) ) = X$.
As a result, the Principle also applies in the reverse direction, from $-d$ to $+d$.

Furthermore, the Principle dovetails neatly with the metaprogram approach to defining defeasible logics \cite{flexf}
in which a $+d$ inference rule is expressed as a logic program defining a predicate $d$,
and the $-d$ inference rule is expressed implicitly as $\mathtt{not}\ d$
(where $\mathtt{not}$ refers to negation as failure).
Thus all logics defined via the metaprogram approach support the Principle\footnote{
However, in one view, defeasible logics WFDL \cite{MG99}, NDL \cite{NuteINAP}, and ADL \cite{MN06}
do not support the Principle.
In the view of \cite{MN10}, these logics have an additional form of failure
(failure-by-loooping) which consequently violates the Principle.
On the other hand, from the point of view of \cite{MG99},
these logics support the Principle but have a more sophisticated evaluation of inference rules \cite{MG99,flexf}.
Despite the differing view on what the inference rules are,
the corresponding proof systems are roughly equivalent (\cite{Maher14}, Theorem 2).
However, none of these logics are characterised by the linear proofs we are using in this paper.
For this reason, we will not consider them further.
Nevertheless, it should be possible to adapt the work in this paper to these logics under the latter view.
}.

However, the advent of $\DL(\pl)$ offers complications to this convenient state of affairs.

\section{Problems}  \label{sect:problems}

The applicability conditions of $\DL(\pl)$
depart from previous logics in the following ways:
\begin{itemize}
\item
They involve proofs/sets ($P_{+\Delta}$ and $P_{+\lambda}$) other than the proof under development.
\item
$-d$ inference rules are not necessary to compute $+d$ inferences.
\item
They use expressions $+d \, q \in X$ in negative contexts\footnote{
A \emph{negative context}, also known as a position of negative polarity,
is a position in a formula which is in the scope of an odd number of negation signs,
assuming formulas composed of $\wedge$, $\vee$, $\neg$, $\forall$, and $\exists$.
A \emph{positive context} is in the scope of an even number of negation signs.
The idea can be extended to other connectives, where it requires more formalization.
}.
\end{itemize}
The first change means that applicability conditions may include reference to other proofs or,
at least, pre-defined sets of conclusions.
The second change implies that negative conclusions might have less importance,
although it is still interesting to know what conclusions cannot be proved.
It conflicts with the common assumption of the presence of negative inference rules,
such as in the Principle of Strong Negation.
The third change introduces a class of expressions that have not previously been used
in defeasible logics.
These changes, particularly the third, bring to light some problems.

One problem is that
coherence can be lost if an inference rule contains $c \notin X$,
if we assume that the Principle of Strong Negation holds.

\begin{example}   \label{ex:incoherent}
Consider the inference rule $+d$
\begin{center}
We may append  $+d \, q$ to $P$ if  $+\Delta \non q  \notin P_\Delta$
\end{center}
This rule expresses a simple form of the Closed World Assumption.
Following the Principle of Strong Negation in \cite{flexf}, the $-d$ inference rule is
\begin{center}
We may append  $-d \, q$ to $P$ if $-\Delta \non q  \notin P_\Delta$
\end{center}
Now consider the defeasible theory consisting only of the single rule
$\non p \rightarrow \non p$.
Then $P_\Delta = \{ -\Delta \, p \}$.
Hence we can infer $+d \, p$ and $-d \, p$.
Thus the logic is incoherent on $D$.
\end{example}

This problem was not previously noticed because defeasible logics had not used non-membership expressions.

In  Appendix C of \cite{sdl}, an ad hoc extension was made to strong negation to address this problem, by defining
\[
\begin{array}{l l l}
sneg(+d{p} \notin X)  & = &  +d{p} \in X \\
sneg(-d{p} \notin X)  & = &  -d{p} \in X \\
\end{array}
\]
However, this means that the definition is now unclear, since we would have, for any proof $X$,
\[
sneg(+d{p} \notin X) = (+d{p} \in X) \ \neq\  (-d{p} \notin X) = sneg(\neg(+d{p} \in X))
\]

So the strong negation of equivalent expressions ($+d{p} \notin X$ and $\neg(+d{p} \in X)$)
can vary, depending on how the expression is written.
This is obviously problematic.

A more serious problem arises from allowing $+d \, q \in X$ in negative contexts
when $X$ is the current proof $P$.
Ideally, 
any inference that can be made at point $i$ in the proof can be deferred to a later point $j$.
That is, if an inference rule is applicable at one point, it should be applicable at all later points.
Similarly, we would wish that any inference that could be made is unaffected
by previously (in the proof) proving other conclusions.
Such inference rules have a kind of stability: their application depends only on
certain triggering criteria having been satisfied.
Stability of inference can be violated if we allow arbitrary use of $+d \, q \in P$ in a negative context.
 
\begin{example}   \label{ex:unstable}
Consider the inference rule
\begin{center}
We may append  $+d \, q$ to $P$ if  $+d \non q  \notin P$
\end{center}
Starting from an empty proof, for any defeasible theory $D$ and literal $q$, we can infer either $+d q$ or $+d \non q$.
However, once (say) $+d q$ is added to the proof, $+d \non q$ can no longer be inferred.

Thus any system attempting to prove a literal faces a nondeterministic choice:
to infer $+d q$ or $+d \non q$.
While this choice can be deferred, and might become moot in the light of other inferences,
it will generally raise the complexity of query-answering and other operations,
as well as the cognitive complexity of understanding the meaning of a theory.
\end{example}

Expanding this idea, it is possible to define an inference rule that treats defeasible rules in defeasible theories
as normal default rules in Reither's default logic \cite{DefaultLogic}.  
The non-determinism in default theories is exhibited by the multiple extensions (in general)
for default theories.
While this could be considered as a sign of the expressiveness of the unrestricted language of applicability conditions, the other side of that coin is cognitive and computational complexity.

Other uses of $P$ also raise problems.

\begin{example}   \label{ex:|P|}
Consider an applicability condition of the form
$(|P| < 5 \rightarrow C_1) \wedge (|P| \geq 5 \rightarrow C_2) $.
This allows a different applicability condition for the first 4 positions in a proof,
than for later positions in a proof.
Such an inference rule can be unstable
(unless $C_1$ and $C_2$ are equivalent).
It makes the first four slots in a proof a kind of limited resource,
which can embed combinatorial issues into the problem of computing proofs.
\end{example}

Instability in inference complicates aspects of proof and consequences:
\begin{itemize}
\item
Although two consequences might be provable separately, they might not be provable in a single proof.
\item
A logic may be inconsistent, even if no proof shows inconsistency.
\item
The notion of a closure becomes messier.  
We cannot rely on a single proof that accumulates all consequences.
There might not be a single minimal set closed under inference.
\item
Appending one proof to another may not result in a proof.
\end{itemize}

Unstable inference rules can produce interesting logics
with nondeterminism in proofs and the ability to treat positions in a proof
as a kind of limited resource.
They introduce a combinatoric aspect to proofs.
But these capabilities seem orthogonal to the main purpose of defeasible logics,
and a distraction from it.
So, although such inference rules might deserve further study,
we will seek to exclude them.

Looked at from another angle,
the syntax $c \notin X$ (and $c \in X$), where $X$ is not the current proof $P$,
suggests the possibility of multiple proofs being constructed simultaneously.
Such an expansion is interesting as providing a distributed model of defeasible reasoning,
as well as a model in which some privacy issues can be addressed
(in a manner similar to distributed constraint satisfaction problems \cite{DCSP}).
However, this also introduces issues of multiple agents, which might have differing motivations,
as well as all the complications that might arise from concurrent computations.
Although this is an intriguing generalization of defeasible logics,
those complications are best addressed after achieving a fuller understanding of defeasible logics
with a single proof.
So we leave logics with multiple simultaneous proofs for another time.

Finally, there was a small problem with the way inference rules were expressed as conditions on proofs.
Consider the inference rule for $\Delta$ expressed as a condition on proofs, 
as in many defeasible logics, since at least \cite{BCN}, 
and as used in original definitions of $\DL(\partial)$ and $\DL(\pl)$:
 
\begin{tabbing}
90123456\=7890\=1234\=5678\=9012\=3456\=\kill

$+\Delta)$  If  $P(i+1) = +\Delta q$  then either \\
\hspace{0.2in}  (1)  $q \in F$;  or \\
\hspace{0.2in}  (2)  $\exists r \in R_{s}[q] \  \forall a \in A(r),
+\Delta a \in P[1..i]$.
\end{tabbing}

The syntax of expressions like $P(i+1)$ and $P[1..i]$
suggests that individual positions in a proof and ranges of elements in a proof are acceptable expressions
in an applicability condition.
However, such expressions can easily lead to instability.

For example, the inference rules
\begin{center}
We may append  $+ d_1 \, q$ to $P$ if $P(1) = +\Delta a$ and ... \\
We may append  $+ d_2 \, q$ to $P$ if $P(1) = +\Delta b$ and ...
\end{center}
lead to nondeterminism in the proofs,
where we can infer $+d_1 p$ only if the first inference in the proof were to infer $+\Delta a$,
and infer $+d_2 p$ only if the first inference in the proof were to infer $+\Delta b$.
As a result, no proof can include both $+d_1 p$ and $+d_2 p$.

Similarly, an inference rule
\begin{center}
We may append  $+ d \, q$ to $P$ if $-\Delta q \in P[1..10]$ and ...
\end{center}
limits the number of $+ d \, q$ inferences to at most ten.
In general, then, a proof must use those first 10 positions to infer sufficient $-\Delta$ conclusions
to allow applications of the $+d$ inference rule.
Different uses of those 10 positions will support different proofs.

\smallskip

Thus we need to carefully define what defeasible inference rules are acceptable.
We will begin this by restricting the syntax of applicability conditions.
We have already modified the syntax to avoid expressions like $P(i)$ and $P[1..i]$.

\section{What is a defeasible logic?}  \label{sect:WDL}

To avoid the problems illustrated above we need to clarify what applicability conditions are permitted,
identify how to avoid non-determinism,
and revise the definition of strong negation.
These issues are addressed in the following three subsections.

\subsection{Language of Applicability Conditions}   \label{sect:LAC}

As a first step, we must specify the language in which applicability conditions are expressed.
One the one hand, we would like as few restrictions as possible, so that we have flexibility
in defining inference rules; on the other hand, we would like to guarantee coherence
while supporting the Principle of Strong Negation.
Previous work on defeasible logics have left the language unaddressed,
except for proposing specific inference rules.

A \emph{proof atomic formula} has the form $+d q \in X$ or $-d q \in X$,
where $d$ is a tag, $q$ is a literal, and $X$ refers to a proof/set
(either $P$ or a pre-defined set of conclusions).
A \emph{pure formula} is a formula not involving a proof atomic formula.
An \emph{applicability condition}
is a first order formula that determines whether an inference may be made.
To address the issue of ambiguity of the ad hoc modification to strong negation in \cite{sdl}, 
we restrict applicability conditions to negation normal form.
A formula is in \emph{negation normal form (NNF)} if all occurrences of $\neg$ are applied to atomic formulas
and the only other connectives are $\wedge$ and $\vee$.
Note that, as a result of the logical connectives permitted, 
formulas occur in a negative context iff they are atomic formulas to which $\neg$ has been applied;
all other formulas are in a positive context.

\begin{definition}   \label{defn:ac}
An \emph{applicability condition} for an inference rule
is a Boolean expression parameterised by three elements:
 the literal $q$ to be inferred;
 the proof $P$ on which it acts;
and
 the defeasible theory $D = (F, R, >)$ from which inferences are made.
In addition, a tag (of the form $+d$ or $-d$) is associated with the inference rule.
 
The syntax of an applicability condition is a logic formula in negation normal form, 
built in the usual way by connectives $\neg$, $\wedge$ and $\vee$ and quantifiers $\exists$ and $\forall$,
from primitive formulas.
Variables that are quantified may range over syntactic elements of $D$, such as rules and literals,
or numbers.
The primitive formulas are:
\begin{itemize}
\item
comparisons involving elements of the defeasible theory $D$ (such as $x \in R_s$ or $s > t$),
\item
arithmetic and set comparisons (such as $|R_s| \geq |R_d|$ or $R[q] \subseteq R_s$), 
\item
membership tests involving the preceding part of the proof (such as $+d' p \in P$), and
\item
membership tests involving tagged literals and pre-defined sets\footnote{
We refer to closures such as $P_\Delta$ and $P_\lambda$ as pre-defined, since
they have been computed from $D$ before the inference rule(s) of interest are applied.
We assume that the inference rules used for any closure are a part of the logic.
}
(such as $+\Delta p \in P_\Delta$),
\end{itemize}
where 
$d'$ is a tag that may be different from $d$.

If, in addition,
$P$ occurs in an applicability condition $C(P)$ only in expressions of the form $\pm d q \in P$ 
and only in a positive context,
then we say $C$ is \emph{$P$-disciplined}.
\end{definition}

The acceptability conditions of inference rules in the logics $\DL(\pl)$, $\DL(\partial)$, 
and other logics in \cite{TOCL10} are in negation normal form, 
since negations are only applied to primitive formulas.
Furthermore, those acceptability conditions are $P$-disciplined.

This definition remains vague as to the terms permitted.
We permit set comprehensions as terms, in order to define notations such as $R[q]$,
but it is not clear what restrictions should be put on such terms.
We exclude the use of $P$ within a set expression, apart from $c \in P$,
so that later distinctions between positive and negative occurrences of expressions $c \in P$
are meaningful\footnote{
Otherwise we could, for example, encode the negative expression $+d q \notin P$ by the positive expression
$+d q \in (\{ +d q ~|~ q \mathrm{~is~a~literal~} \} \backslash P)$.
}, but further restrictions might be desirable.
Similarly, it is not clear what restrictions should be placed on the use of arithmetic functions.
Certainly complexity, and even computability, can be affected by (the lack of) restrictions,
but these considerations are left for another time.

Finally, limitations on what pre-defined sets should be allowed in an applicability condition
also require further thought.
In $\DL(\pl)$, $P_\lambda$ and $P_\Delta$ are stratified in the sense that,
while there is a dependency of $P_\lambda$ on $P_\Delta$,
there is no dependency in the reverse direction.
More importantly, neither set depends on $\pl$ inference rules.
(This stratification appears in the metaprogram representation of $\DL(\pl)$ 
as a stratification of the corresponding logic program \cite{sdl2}.)
While stratification provides clarity of meaning for the inference rules,
non-stratified inference might permit a form of reflection in the logics.
For definiteness and simplicity, we assume that pre-defined sets are stratified and, in particular,
do not depend on the tag being defined.

An \emph{inference rule} has the form
\begin{center}
We may append  $\pm  d \, q$ to $P$ if $C$
\end{center}
where $C$ is an applicability condition,
$\pm$ is either $+$ or $-$,
and $\pm d$ is the tag associated with the inference rule.
All conclusions inferred by this inference rule have this tag.
Also, no two inference rules may have the same tag.
(This imposes little weakening of expressive power because
multiple inference rules for the same tag can be replaced by a single inference rule
whose applicability condition is the disjunction of the applicability conditions\footnote{
There is a minor weakening in that applicability conditions $U \wedge C$ and $\neg U \wedge C$
would be replaced essentially by $C$, assuming the use of classical logic.
If $U$ is unknown, uncomputable or imponderable in some way
(say, $U$ is $P \neq NP$ or ``the Axiom of Choice holds'')
then the result is arguably different from what was intended.
}.)

Formulas are restricted to negation normal form with only the basic connectives 
to avoid the ambiguity in \cite{sdl} and so that
the notions of positive and negative context are simple.
However, we could dispense with negation normal form, 
at the expense of complicating these and other definitions.

We will need the following definitions later.
An applicability condition $C$ is said to be \emph{pos-only} if 
every proof atomic formula in $C$ appears only in a positive context.
Similarly, $C$ is \emph{neg-only} if proof atomic formulas appear only in a negative context.
It is easy to see that if $C$ is pos-only then $\neg C$ is neg-only, and vice versa.
Before $\DL(\pl)$, all defeasible logic inference rules were pos-only. 

$P$-discipline excludes problematic logics such as those in Examples~\ref{ex:unstable} and \ref{ex:|P|}.
It overlaps with pos-only formulas, but they are distinct:
$P$-discipline does not address pre-defined sets, while pos-only does not address uses of $P$
other than in proof atomic formulas.
Any inference rule $d$ with a $P$-disciplined applicability condition,
has the property that if $d$ is applicable to proof $Q$ then $d$ is applicable to any permutation of $Q$.

We can now define an umbrella notion of defeasible logic that admits both
$\DL(\partial)$ and $\DL(\pl)$.
A \emph{defeasible logic} consists of 
a language in which a defeasible theory $D$ may be written and
a finite set of inference rules
(each with an applicability condition satisfying Definition~\ref{defn:ac}).
Each inference  rule is
labelled by a tag that also appears with each conclusion drawn by the inference rule.
Tags may have the form $+d$, indicating provability, or $-d$, indicating provable non-provability.
A \emph{propositional defeasible logic} is a defeasible logic that restricts defeasible theories
to use a finite set of literals.
A \emph{proof} $P$ is a sequence of conclusions, each drawn from $D$ by an inference rule applied to preceding conclusions in $P$.

This definition of defeasible logics sets the bounds of what might be a defeasible logic,
but does not address the main problems identified in Section~\ref{sect:problems}.
We now turn to the problem of instability.

\subsection{Stability}

As indicated earlier, we want to restrict our attention to defeasible logics that are stable.
First we need to formalize the notion of stability.

We formulate stability as follows\footnote{
This notion of stability is different from stability addressed in \cite{Billington93},
which is a generalisation of the notion of atomic stability of nonmonotonic inheritance networks \cite{HTT90}.
That notion of stability concerns the stability of the consequences of a defeasible theory $D$ when
a provable literal is added to $D$, a form of cautious monotonicity \cite{KLM}.
In comparison,
the stability defined here is stability of applicability of inference rules when
provable (tagged) literals are added to the proof.
}.
Recall that $P$ is a parameter to $C$, so that we can write $C(P)$.

\begin{definition}
An inference rule
\begin{center}
We may append  $\pm d \, q$ to $P$ if $C(P)$
\end{center}
is \emph{stable}
if, for any proof $P$ and any proof $Q$ containing $P$ as a subsequence\footnote{
A sequence $s$ is a \emph{subsequence} of $a_1 \ldots a_n$ if
$s$ is empty or
has the form $a_{i_1} \ldots a_{i_k}$ for some $k \leq n$ where
$1 \leq i_1 < i_2 < \cdots < i_k \leq n$.
}, $C(P) \rightarrow C(Q)$.

A defeasible logic is stable if all inference rules in the logic are stable.
\end{definition}

A weaker version of stability, which still addresses the ability
to defer application of an inference rule, is the following.
\begin{definition}
An inference rule
\begin{center}
We may append  $\pm d \, q$ to $P$ if $C(P)$
\end{center}
is \emph{applicability persistent}
if, for any proof $P$ and any proof $Q$ containing $P$ as a prefix, $C(P) \rightarrow C(Q)$.

A defeasible logic is applicability persistent if all inference rules in the logic are applicability persistent.
\end{definition}

Clearly, every stable inference rule is applicability persistent.
They are both versions of monotonicity for proofs, rather than sets.
In general, however, they are not equivalent.

\begin{proposition}
There is an inference rule that is applicability persistent but not stable.
\end{proposition}
\skipit{
}
\begin{proof}
Consider a defeasible logic $\DL(d)$ that contains the inference rules of $\DL(\partial)$ 
(i.e. $+\Delta$, $-\Delta$, $+\partial$ and $-\partial$)
and the following inference rule
\begin{center}
\begin{tabular}{rl}
We may append  $+ d \, q$ to $P$ & if $P(1) = +\Delta a$ and $C_\partial (P)$ \\
& or $P(1) \neq +\Delta a$ and $C_\Delta (P)$
\end{tabular}
\end{center}
where $a$ is a literal in the language of the logic,
$C_\partial$ is the applicability condition for $+\partial$, and
$C_\Delta$ is the applicability condition for $+\Delta$.

Note that every proof that starts with $+\Delta a$ is a proof in $\DL(\partial)$, 
if we replace $+d$ by $+\partial$.
Similarly, every proof that does not start with $+\Delta a$ is a proof in $\DL(\Delta)$, 
if we replace $+d$ by $+\Delta$.
Both $\DL(\partial)$ and $\DL(\Delta)$ are applicability persistent, 
by Proposition~\ref{prop:stable} below.
Consequently, the defeasible logic under consideration is applicability persistent.

On the other hand, the defeasible logic is not stable.
Consider the defeasible theory $D = (\{a, b\}, \{ a \Rightarrow c \}, \emptyset)$,
the proof $P = +\Delta a ; +d a$ and the proof $Q = +\Delta b ; +\Delta a ; +d a$.
Clearly $P$ is a subsequence of $Q$.
$C(c, P)$ holds, because $C_\partial(c, P)$ holds.
$C(c, Q)$ does not hold, because $C_\Delta(c, Q)$ does not hold.
Thus $C(c, P) \rightarrow C(c, Q)$ is not valid.
Hence, the logic is not stable.
\end{proof}

The inference rule used to create the distinction demonstrated above relies on a syntax that we have already discarded.
The construction of this distinction involves the combination of two inference rules,
with an element of the proof causing a switch from the stronger to the weaker inference rule.
The insertion of this element in $Q$ activates instability, but if the element is established in the middle
of a proof then it can also violate applicability persistence.
So this approach to identifying a distinction seems viable only if we can identify the initial conclusion of a proof.
Given the restriction of syntax we have already imposed, it is not clear that there is a practical distinction
between the two concepts.

Furthermore,
restricting the use of $P$ in applicability conditions can achieve stability and applicability persistence.
It thus renders the distinction largely moot,
although there might be a weaker restriction that assures applicability persistence but not stability.

\begin{proposition}   \label{prop:stable}
Consider an inference rule
\begin{center}
We may append  $+ d \, q$ to $P$ if $C$
\end{center}
If 
$C$ is $P$-disciplined
then the inference rule is stable (and applicability persistent).
\end{proposition}
\begin{proof}
Let $P'$ be a proof that is a subsequence of $Q'$.
For any atomic formula $a$ of $C$ not involving $P$, $a$'s evaluation is the same in $P'$ and $Q'$.
If $\pm d q \in P'$ then also $\pm d q \in Q'$, since $P$ is a subsequence of $Q$.
Consequently, since such formulas only occur in $C$ in a positive context,
for any literal $q$ and defeasible theory $D$, if $C(P')$ holds then so too does $C(Q')$.
\end{proof}

Consequently, by inspection of their inference rules,
both $\DL(\partial)$ and $\DL(\pl)$ are stable, as are the logics in \cite{TOCL10}.

Stability is sufficient to do common kinds of manipulations of proofs,
and to ensure that the idea of closure is well-defined.

\begin{proposition}   \label{prop:stable2}
Consider a stable defeasible logic, a subset $J$ of the inference rules, and a defeasible theory $D$.

\begin{enumerate}
\item
If $P$ is a proof, any prefix of $P$ is a proof.
\item
If $P$ and $Q$ are proofs then their concatenation $P; Q$ is a proof.
\item
If $P$ and $Q$ are proofs then any interleaving of $P$ and $Q$ is a proof.
\item
The $J$-closure $P_J$ of the $J$ inference rules on $D$ is well-defined.
\end{enumerate}

\end{proposition}
\skipit{
\begin{proof}
1.
This follows immediately from the definition of proof, independent of stability.

2. 
Let $Q_i$ be any prefix of $Q$.  Let $\pm d_i q_i$ be the next conclusion in $Q$, and
let $C_i$ be the corresponding applicability condition.
Then $C_i(q_i, Q_i)$ is true.  
By stability, it follows that $C_i(q_i, P;Q_i)$ is true.
Hence each inference rule used in $P; Q$ is applicable when it is used.
That is, $P; Q$ is a proof.

3. 
Let $S$ be an interleaving of $P$ and $Q$.
Let $\pm d_i q_i$ be the $i^{th}$ conclusion in $S$ and
let $C_i$ be the corresponding applicability condition.
Let $P_i$ and $Q_i$ be the prefixes of $P$ and $Q$ respectively that appear before position $i$ in $S$,
and let $S_i$ be the prefix of $S$ before position $i$.
If the $i^{th}$ conclusion occurs in $P$ then $C_i(q_i, P_i)$ is true;
otherwise, $C_i(q_i, Q_i)$ is true.
By stability, $C_i(q_i, S_i)$ is true.
Hence each inference rule used in $S$ is applicable when it is used.
That is, $S$ is a proof.

4.  
Let $I$ be the smallest reference-closed set containing $J$.
Given $D$, each $I$-consequence has a proof that establishes it.
Concatenating all these proofs creates (by part 2) a single proof $Q_I$ that contains all $I$-consequences.
$Q_I$ is finite because the number of tags and number of literals is finite.
Considered as a set,
$Q_I$ is $I$-closed because, by its definition, it contains all $I$-consequences.

Suppose $Q_I$ is not the smallest $I$-closed set.
Then there is some $S$ that is $I$-closed and $Q_I \backslash S \neq \emptyset$.
Let $c \in Q_I \backslash S$ and consider a proof $P_c$ for $c$.
Let $c'$ be the first element in $P_c$ that is not in $S$.
Then $c'$ can be inferred from a sequence from $S$ but $c' \notin S$.
Thus $S$ is not $I$-closed.
Hence $Q_I$ is the $I$-inferential closure.

Consequently, the definition of $J$-closure based on concatenated proofs and 
the definition based on closed sets are equal, and the notion of  $J$-closure is well-defined.
\end{proof}
}

Note that the proof constructed in part 4 is not duplicate-free,
as was required in defining $P_\Delta$ in Section~\ref{sect:sdl}.
Ideally, we would like to be able to delete duplicate conclusions from a proof.
However, the class of logics under consideration includes those where such an operation is unsound;
the resulting sequence is not a proof.

\begin{example}   \label{ex:delete}
Consider a defeasible logic $\DL(\partial')$, like $\DL(\partial)$ but with a slightly altered applicability condition:
instead of requiring (in $+\partial (2.1)$) each literal $a$ in the antecedent of a rule to be proved in $P$,
$+\partial'$ requires that there is a separate occurrence of $+\partial' a$ for each $a$ in the antecedent.
$\DL(\partial')$ is stable.

In that case, if $D$ consists of a rule $a, a \Rightarrow b$ and a fact $a$ then
a duplicate occurrence of $+\partial' a$ is necessary to infer $+\partial' b$ using this rule.
Thus, while $+\Delta a; +\partial' a; +\partial' a; +\partial' b$ is a valid proof,
the deletion of a duplicate results in $+\Delta a; +\partial' a; +\partial' b$,
which is not a proof.
\end{example}

The next example considers a different modification of $\DL(\partial)$,
one that corresponds to a memory-saving choice to discard ``old'' conclusions from a proof.

\begin{example}  \label{ex:recent}
Consider a defeasible logic $\DL(\partial')$ like $\DL(\partial)$ except that
references to $c \in P$ in $+\partial$ and $-\partial$ inference rules are replaced by $c \in P[\floor{\frac{i}{2}}..i]$.
That is, only ``recent'' conclusions may be used to support an inference, where recent refers to the latter half of  $P$.
This logic is not stable, not even applicability persistent.
Any recent conclusion $c$ that supports an inference from $P$ can become old (i.e. not recent) by the insertion of sufficiently many unrelated conclusions after $c$.
Thus $C'(P)$ holds but $C'(Q)$ does not
(where $C'$ is the applicability condition for $+\partial'$ and $Q$ is a proof that has $P$ as a subsequence,
but has inserted the unrelated conclusions).

Obviously, since $\partial'$ is a restriction of $\partial$ that looks at only half of $P$ 
and the inference rules are pos-only,
any proof of $\DL(\partial')$ is also a proof of $\DL(\partial)$.

Nevertheless, $\DL(\partial)$ and $\DL(\partial')$ have the same consequences.
Consider any defeasible theory $D$ and consequence $c$ of $D$ in $\DL(\partial)$.
Let $N$ be the length of a proof $P$ of $c$,
and let $c_1$ be the first conclusion in $P$.
Now consider the sequence consisting of $N$ copies of $c_1$ followed by $P$.
This is a proof, and all conclusions in $P$ are recent.
Thus $c$ is a consequence of $\DL(\partial')$.

This shows that $\DL(\partial)$ and  $\DL(\partial')$
are equivalent in terms of consequences.
\end{example}

In the above example we see an unstable logic that is equivalent, with respect to the consequences it infers,
to a stable logic.
This emphasises that stability is a property of proofs and inference rules, rather than
consequences.

The work in this section has resolved many of the problems discussed in Section~\ref{sect:problems},
but we still need to adapt the definition of strong negation.

\subsection{Revision of Strong Negation}  \label{sect:SN2}
  
With the clarification of applicability conditions,
we can present a revised definition of strong negation.
$sneg$ is defined inductively on the set of negation normal form formulas.
Let $d$ be any proof tag, $X$ be any sequence or set, and $A$ and $B$ be any formulas.
\[
\begin{array}{l l l}
sneg(+d{p} \in X)  & = &  -d{p} \in X \\
sneg(-d{p} \in X)  & = &  +d{p} \in X \\
sneg(+d{p} \notin X)  & = &  +d{p} \in X \\
sneg(-d{p} \notin X)  & = &  -d{p} \in X \\
sneg( A \wedge B )  & = &  sneg( A ) \vee sneg( B ) \\
sneg( A \vee B )  & = &  sneg( A ) \wedge sneg( B ) \\
sneg( \exists x ~ A )  & = &  \forall x ~ sneg( A )  \\
sneg( \forall x ~ A )  & = &  \exists x ~ sneg( A )  \\
sneg( \neg A )  & = &  \phantom{\neg} A    \hspace{1.1in} \mbox{if $A$ is a pure atomic formula} \\
sneg( A )  & = &  \neg A                             \hspace{1.1in} \mbox{if $A$ is a pure atomic formula} \\
\end{array}
\]
Recall that a \emph{pure formula} is a formula that does not contain a tagged literal.
This definition refines the rather ad hoc modification in \cite{sdl}.

The main differences from the definition in \cite{flexf} are that
pre-defined sets are accommodated,
non-membership tests (like $+d{p} \notin X$) are defined directly,
while negation is only addressed when it is applied to atomic formulas.
The treatment of non-membership avoids the problem discussed in Example~\ref{ex:incoherent},
while the treatment of negation avoids ambiguity.

Under the old definition, $sneg$ was a self-inverse,
that is, $sneg( sneg( C ) ) = C$.
That is no longer the case because
$sneg( sneg( +d{p} \notin X ) ) = sneg( +d{p} \in X ) = (-d{p} \in X)$.
Consequently, we no longer have the property under the Principle of Strong Negation that 
if $C$ is the applicability condition for $-d$ then $sneg(C)$ is the applicability condition for $+d$. 
Indeed, $sneg$ no longer has an inverse,
because both $-d{p} \in X$ and $+d{p} \notin X$ are mapped to $+d{p} \in X$.
Furthermore, this definition loses the close fit between strong negation and
negation-as-failure in the metaprogramming approach.

It is worth noting that if $C$ is $P$-disciplined then $sneg(C)$ is $P$-disciplined too
(and consequently stable, by Proposition~\ref{prop:stable}).
Similarly, it is easy to see from the definition of $sneg$ that $sneg(C)$ is always pos-only.

If there are no proof atomic formulas in negative contexts, no pre-defined sets,
and $C$ is in negation normal form
then this definition is equivalent to that of \cite{flexf}.
To this extent, the new definition is a conservative extension of the original definition.
Consequently, under those circumstances $sneg$ is a self-inverse.
More generally,

\begin{proposition}  \label{prop:selfinverse}
If an applicability condition $C$ is pos-only
then $sneg$ is a self-inverse.
That is, 
$sneg(sneg(C)) = C$,
or $sneg^2 = 1$.
\end{proposition}

Proof of this proposition and of the next three are straightforward inductions on the structure of $C$.

On all applicability conditions this form of strong negation satisfies the following weaker property.
\begin{proposition}
For every applicability condition $C$, $sneg(sneg(sneg(C))) = sneg(C)$,
or $sneg^3 = sneg$.

Consequently, $sneg^{2n+1} = sneg$ and $sneg^{2n} = sneg^2$ for $n = 1,2,\ldots$.
\end{proposition}

An advantage of the revised definition is 
that, for logics like those in \cite{TOCL10}, strong negation is a left inverse of negation.
In this proposition we assume that negated formulas are converted into NNF.

\begin{proposition}   \label{prop:inverse}
\noindent
\begin{enumerate}
\item
If an applicability condition $C$  is pos-only
then $sneg( \neg C )$ is identical to $C$.
\item
If an applicability condition $C$ is neg-only
then $sneg( C )$ is identical to $\neg C$.
\item
For any applicability condition $C$, $sneg(C)$ is pos-only.
\end{enumerate}
\end{proposition}

We distinguish two notions of satisfiability for applicability conditions in a logic $\DL(d)$.
There are three free variables in applicability conditions: defeasible theories, literals, and sequences of conclusions.
We say a formula $\psi$ is \emph{proof-satisfiable} if
there is a defeasible theory $D$, a literal $q$, and a proof $P$ in $\DL(d)$
such that $\psi(D, q, P)$ is valid.
$\psi$ is \emph{seq-satisfiable} if
there is a defeasible theory $D$, a literal $q$, and a sequence $P$ of conclusions in $\DL(d)$
such that $\psi(D, q, P)$ is valid.
Obviously, if a formula is proof-satisfiable then it is also seq-satisfiable.
Correspondingly, there is proof-validity and seq-validity,
and if a formula is seq-valid then it is also proof-valid.
Almost always we use proof-satisfiability and -validity.

The following proposition justifies the name ``strong negation''.
\begin{proposition}  \label{prop:strong}
Consider a defeasible logic where all tags and pre-defined sets are coherent.
For every applicability condition $C$

$sneg(C) \rightarrow \neg C$ is proof-valid.
\end{proposition}

\smallskip
The statement of the Principle of Strong Negation remains the same but,
because of the definition of applicability condition and the revised definition of $sneg$,
its effect is slightly different.
\smallskip

Although the revised definition of strong negation fixes some problems,
the introduction of pre-defined sets undermines the Principle of Strong Negation,
in the sense that pre-defined sets are not subject to the Principle.

\begin{example}   \label{ex:PoSNsets}
We compare the effect of different forms of closure.
Let $P_{+\Delta}$ denote the closure of the $+\Delta$ inference rule,
and $P_\Delta$ denote the closure of the $+\Delta$ and $-\Delta$ inference rules.
Let $+d_1$ be
\begin{center}
We may append  $+d_1 \, q$ to $P$ if $+\Delta q \in P_{+\Delta}$
\end{center}
and $+d_2$ be
\begin{center}
We may append  $+d_2 \, q$ to $P$ if $+\Delta q \in P_{\Delta}$
\end{center}
Clearly the inference rules $+d_1$ and $+d_2$ are equivalent.

Applying the Principle of Strong Negation, 
$-d_2$ is
\begin{center}
We may append  $-d_2 \, q$ to $P$ if $-\Delta q \in P_{\Delta}$
\end{center}
and
$-d_1$ is
\begin{center}
We may append  $-d_1 \, q$ to $P$ if $-\Delta q \in P_{+\Delta}$
\end{center}
But, since $P_{+\Delta}$ contains only $+\Delta$ conclusions, $-d_1$ is equivalent to
\begin{center}
We may append  $-d_1 \, q$ to $P$ if $false$
\end{center}
\end{example}

Thus, although the letter of the Principle is supported by $\DL(d_1)$,
the spirit is violated:
provable failures of $+d_1$ that might be detected are not, because $P_{+\Delta}$ 
does not contain information about provable failures of $+\Delta$.
If we insist that closures are closed under both $+$ and $-$ tags, for each tag present in the closure,
then the example above is rectified;
the $\Delta$ closure is required to have the ``missing'' $-\Delta$ conclusions.
In terms of $J$-closure in Definition~\ref{defn:closure}, we require that $J$ consists of matching tags.
In such cases we say that the logic has \emph{even-handed} closures.
However, in general, pre-defined sets are permitted to contain \emph{any} collection of conclusions,
so their presence remains a problem.

We now turn to the effect that this revised Principle has on coherence.
First, we notice that the revision fixes the problem in Example~\ref{ex:incoherent}.

\begin{example}   \label{ex:incoherent2}
Recall that, in Example~\ref{ex:incoherent},
the original Principle of Strong Negation applied to the inference rule
\begin{center}
We may append  $+d \, q$ to $P$ if  $+\Delta \non q  \notin P_\Delta$
\end{center}
led to incoherence
on the defeasible theory consisting of the single rule $\non p \rightarrow \non p$.

Applying the revised Principle of Strong Negation, the $-d$ inference rule becomes
\begin{center}
We may append  $-d \, q$ to $P$ if $+\Delta \non q  \in P_\Delta$
\end{center}
With this definition, we cannot infer $-d \, p$ and 
the logic consisting of these inference rules (and the $\Delta$ inference rules) is coherent on $D$.
More generally, the logic is coherent
(see Theorem~\ref{thm:cohere}).
\end{example}

Whether or not the Principle of Strong Negation is supported,
coherence is closely related to the joint satisfiability of applicability conditions.
\begin{proposition}  \label{prop:sneg}
Consider a stable defeasible logic with tag $d$.
Let the $+d$ inference rule be
\begin{center}
We may append  $+d \, q$ to $P$ if $C^+$
\end{center}
and the $-d$ inference rule be
\begin{center}
We may append  $-d \, q$ to $P$ if $C^-$
\end{center}
Then
$C^+ \wedge C^-$ is proof-unsatisfiable iff $d$ is coherent
\end{proposition}
\skipit{
\begin{proof}
Suppose $d$ is incoherent.
Then there is a proof $P$ containing both $+d q$ and $-d q$, for some $D$ and $q$.
At the end of $P$, by stability, both $+d$ and $-d$ can be applied.
Hence, both $C^+(q, P)$ and $C^-(q, P)$ are satisfied.
That is, $C^+ \wedge C^-$ is proof-satisfiable.

Suppose $C^+ \wedge C^-$ is proof-satisfiable.
Then there is a literal $q$ and proof $P$ that satisfy $C^+ \wedge C^-$.
That is, both $+d q$ and $-d q$ can be inferred.
By stability, after one inference is made the other remains applicable.
Thus $P; +d q; -d q$ is a proof that demonstrates that $d$ is incoherent.
\end{proof}
}

Applying this result to logics supporting the Principle of Strong Negation, we have
\begin{corollary}  \label{cor:coherent}
Consider a stable defeasible logic with tag $d$.
Let the $+d$ inference rule be
\begin{center}
We may append  $+d \, q$ to $P$ if $C$
\end{center}
If the logic supports the Principle of Strong Negation
then

$C \wedge sneg(C)$ is proof-unsatisfiable iff $d$ is coherent
\end{corollary}
\skipit{
\begin{proof}
By the previous proposition, with $C$ for $C^+$ and $sneg(C)$ for $C^-$.
\end{proof}
}

We can formulate a sufficient condition for consistency
in terms of the applicability condition.
\begin{proposition}   \label{prop:cons}
Consider a stable defeasible logic.
Let $C_d$ be the applicability condition for an inference rule $+d$
and $C_\Delta$ be the applicability condition for $+\Delta$.

If
$~ \forall D \forall P \forall q ~ ~(C(q) \wedge C(\non q))~ \rightarrow~  (C_\Delta(q) \wedge C_\Delta(\non q)) ~$
is proof-valid
then the inference rule $+d$ is consistent.

\end{proposition}
\skipit{
\begin{proof}
Let $D$ be a defeasible theory and $q$ be a literal.
Suppose $+d q$ and $+d \non q$ are consequences.
Consider the closure $P_d$ closed under $\pm d$ and $\pm \Delta$,
which exists, by Proposition~\ref{prop:stable2}.
Then $(C_d(q, P_d) \wedge C_d(\non q, P_d))$ is proof-valid.
Consequently, $(C_\Delta(q, P_d) \wedge C_\Delta(\non q, P_d)) $ is proof-valid.
Since $P_d$ is closed, it follows that $+\Delta q \in P_d$ and $+\Delta \non q \in P_d$.
That is $+\Delta q$ and $+\Delta \non q$ are consequences.
This holds for any $D$ and any $q$.
Thus the inference rule $+d$ is consistent.
\end{proof}
}

It seems unlikely that the sufficient condition is necessary,
because that condition requires that \emph{any} proof $P$ able to infer conflicting $+d$-consequences is also 
immediately able to infer the corresponding conflicting $+\Delta$-consequences, 
whereas consistency only requires that  \emph{there exists} a proof with the conflicting $+\Delta$-consequences.
Nevertheless, it remains to find a logic that exhibits this distinction.

The last three results provide a basis for testing for coherence and consistency
in stable defeasible logics.
While reasoning about proof-validity of first-order formulas might be challenging,
it is still easier than reasoning directly about proofs.

\section{Well-behaved Defeasible Logics}  \label{sect:WBDL}

The criterion we suggest for a ``true'' defeasible logic is as follows.

\begin{definition}
We say a defeasible logic is \emph{well-behaved}
if it is stable, supports the revised Principle of Strong Negation, and
all pre-defined sets in inference rules are coherent even-handed closures.
\end{definition}

$\DL(\partial)$ is well-behaved because all its inference rules are $P$-disciplined, and hence stable,
it satisfies the original Principle of Strong Negation and has no use of non-membership
(so it also satisfies the revised Principle),
and uses no pre-defined sets.
The same applies to other logics defined in \cite{TOCL10}.

$\DL(\pl)$  as defined in Section~\ref{sect:sdl} and the body of \cite{sdl}
might not be considered well-behaved, but this is only a technicality.
If we replace uses of $P_{+\Delta}$ and $P_{+\lambda}$ by $P_\Delta$ and $P_\lambda$ (respectively)
and add inference rules for
$-\Delta$, $-\lambda$ and $-\pl$ as defined in Appendix C of \cite{sdl}
(which are in accordance with the Principle of Strong Negation),
then the resulting logic has exactly the same $+$-consequences as $\DL(\pl)$
(and exactly the same proofs involving only $+$-consequences).
This logic is stable (by Proposition~\ref{prop:stable}),
supports the revised Principle of Strong Negation,
and the only pre-defined sets used are coherent even-handed closures.
Thus it is well-behaved.
Both logics do not present applicability conditions in the syntax required by Definition~\ref{defn:ac},
but can easily be rewritten in that form.

We can now establish that well-behaviour is a sufficient condition for a defeasible logic to be coherent.
Versions of the following theorem were given in \cite{flexf,DL_intent}.
The result in \cite{flexf} applied to defeasible logics formulated by metaprogram.
\cite{DL_intent} does not specify the class of defeasible logics and its proof is rather vague.
Both apply to a narrower class of inference rules;
neither address the problem of incoherence identified in Example~\ref{ex:incoherent}. 

\begin{theorem}  \label{thm:cohere}
Every well-behaved defeasible logic is coherent.
\end{theorem}
\skipit{
\begin{proof}
Suppose, to obtain a contradiction, 
that a well-behaved logic is not coherent.
By stability, and Proposition~\ref{prop:stable2}, any incoherence can be shown in a single proof.
Let $D$ be a defeasible theory, $d$ a tag and 
$q$ be a literal that demonstrate non-coherence by a single minimal length proof.
That is,
let $P'$ be a minimal length proof such that each of $+d q$ and $-d q$ can be inferred from $P'$ in one inference step. 

Let the $+d$ inference rule have the form
\begin{center}
We may append  $+d \, q$ to $P$ if  $C_d$
\end{center}
Then the $-d$ inference rule is
\begin{center}
We may append  $-d \, q$ to $P$ if  $sneg(C_d)$
\end{center}

We claim that, for any subexpression $C$ of $C_d$, $C \wedge sneg(C)$ is proof-unsatisfiable\footnote{
For brevity, we simply use ``unsatisfiable'' within this proof.
}.
The proof is by induction on the structure of $C_d$.

If $C$ is $+d' p \in P$ (or $-d' p \in P$),
where $d'$ is a tag possibly different from $d$ and $p$ is possibly different from $q$,
then $sneg(C)$ is $-d' p \in P$ (or $+d' p \in P$)
and $C \wedge sneg(C)$ is $+d' p \in P \wedge -d' p \in P$.
Since we assumed that $P'$ is a minimal length proof for non-coherence,
$P'$ does not contain an incoherent tagged literal and hence one of the conjuncts is false.
Thus $C \wedge sneg(C)$ is unsatisfiable on $P'$.

If $C$ is $+d' p \notin P$ (or $-d' p \notin P$)
then $sneg(C)$ is $+d' p \in P$ (or $-d' p \in P$)
so $C \wedge sneg(C)$ is unsatisfiable.

If $C$ is $+d' p \in X$ (or $-d' p \in X$) where $X$ is pre-defined,
then $sneg(C)$ is $-d' p \in X$ (or $+d' p \in X$)
and $C \wedge sneg(C)$ is unsatisfiable,
because $X$ is coherent.

If $C$ is $+d' p \notin X$ (or $-d' p \notin X$)
then $sneg(C)$ is $+d' p \in X$ (or $-d' p \in X$)
so $C \wedge sneg(C)$ is unsatisfiable.

If $C$ is pure then $sneg(C)$ is $\neg C$.
Thus $C \wedge sneg(C)$ is unsatisfiable.

If $C$ is $A \wedge B$
then $sneg(C)$ is $sneg(A) \vee sneg(B)$,
and $C \wedge sneg(C)$ is $A \wedge B \wedge (sneg(A) \vee sneg(B))$.
By the induction hypothesis,
both $A \wedge sneg(A)$ and $B \wedge sneg(B)$ are unsatisfiable.
Hence $C \wedge sneg(C)$ is unsatisfiable.

If $C$ is $A \vee B$
then $sneg(C)$ is $sneg(A) \wedge sneg(B)$,
and $C \wedge sneg(C)$ is $(A \vee B) \wedge (sneg(A) \wedge sneg(B))$.
By the induction hypothesis,
both $A \wedge sneg(A)$ and $B \wedge sneg(B)$ are unsatisfiable.
Hence $C \wedge sneg(C)$ is unsatisfiable.

If $C$ is $\exists x ~ A$
then $sneg(C)$ is $\forall x ~ sneg(A)$.
Let $A\theta$ be an instance of $A$ demonstrating $\exists x ~ A$.
By the induction hypothesis,
$A \wedge sneg(A)$  is unsatisfiable.
Hence $A\theta \wedge sneg(A)\theta$  is unsatisfiable.
It follows that $\forall x ~ sneg(A)$  is unsatisfiable, and so
$\exists x ~ A \wedge \forall x ~ sneg(A)$ is unsatisfiable.

If $C$ is $\forall x ~ A$
then $sneg(C)$ is $\exists x ~ sneg(A)$.
Let $sneg(A)\theta$ be an instance of $sneg(A)$ demonstrating $\exists x ~ sneg(A)$.
By the induction hypothesis,
$A \wedge sneg(A)$  is unsatisfiable.
Hence $A\theta \wedge sneg(A)\theta$  is unsatisfiable.
It follows that $\forall x ~ A $  is unsatisfiable, and so
$\forall x ~ A \wedge \exists x ~ sneg(A)$ is unsatisfiable.

Thus, by induction, $C \wedge sneg(C)$ is unsatisfiable on $P'$, for any subexpression $C$ of $C_d$.
In particular, $C_d \wedge sneg(C_d)$ is unsatisfiable on $P'$.
But this contradicts the original assumption that both $+d q$ and $+d q$ are applicable on $P'$.
Hence there is no such $q$ and $P'$,
and thus
$d$ is coherent.
Since this holds for any $d$, the logic is coherent.
\end{proof}
}

We can restate this theorem in terms of applicability conditions.

\begin{corollary}   \label{cor:unsat}
Let $C$ be the applicability condition for an inference rule in a well-behaved defeasible logic.

Then $C \wedge sneg(C)$ is proof-unsatisfiable.
\end{corollary}
\skipit{
\begin{proof}
By the previous theorem, the logic is coherent.
By Corollary~\ref{cor:coherent}, $C \wedge sneg(C)$ is proof-unsatisfiable.
\end{proof}
}

We also can see that a double strong-negation does not result in any weakening.

\begin{corollary}  \label{cor:s^2}
Let $C$ be an applicability condition for an inference rule in a well-behaved defeasible logic.

Then
$sneg(sneg(C)) \rightarrow C$ is proof-valid.
\end{corollary}
\skipit{
\begin{proof}
By the previous theorem, the logic is coherent.
Furthermore, every pre-defined set is coherent, from the definition of well-behaved.
Thus $+d q \in X \rightarrow -d q \notin X$ and $-d q \in X \rightarrow +d q \notin X$ are proof-valid,
where $X$ is either $P$ or a pre-defined set.
The proof is now a straightforward induction on the structure of $C$.
\end{proof}
}

Theorem \ref{thm:cohere} does not need even-handedness, but it does require each of the other elements of well-behaviour.
Fairly obviously,
the coherence of pre-defined sets is necessary.  Consider the inference rule
\begin{center}
We may append  $+d \, q$ to $P$ if  $+d' q \in X$
\end{center}
which simply copies positive conclusions from $X$,
and similarly for $-d$, as required by the Principle of Strong Negation.
Any incoherence in $d'$-consequences in $X$ will be reflected in the $d$-consequences.

Also obviously, without the Principle $+d$ and $-d$ can be independent of each other, 
allowing an example as simple as
\begin{center}
We may append  $+d \, q$ to $P$ if  $true$ \\
We may append  $-d \, q$ to $P$ if  $true$
\end{center}

It might be wondered if stability is used only to simplify the proof,
since its main use is to accumulate the incoherence in a single proof.
However, we will see that it also is needed.

\begin{example}
Consider a defeasible logic $\DL(d)$ with the monotonic inference rules $+\Delta$ and $-\Delta$
and defeasible rules $+d$ and $-d$.
The $+d$ inference rule is
\begin{center}
\begin{tabular}{rl}
We may append  $+ d \, q$ to $P$ & if $+\Delta a \in P$ and $q \in F$ or \\
& if $+\Delta a \notin P$ and $q \notin F$
\end{tabular}
\end{center}
where $a$ is a literal in the language of the logic and
$C_\Delta$ is the applicability condition for $+\Delta$.

It is clear this logic is not stable: an inference $+d q$, for some fact $q$,  that is valid from the empty proof
is not valid from the proof consisting only of $+\Delta a$  (assuming a defeasible theory containing facts $a$ and $q$).

Following the Principle of Strong Negation, the inference rule for $-d$ is
\begin{center}
\begin{tabular}{rl}
We may append  $- d \, q$ to $P$ & if $+\Delta a \in P$ and $q \notin F$ or \\
& if $-\Delta a \in P$ and $q \in F$
\end{tabular}
\end{center}
obtained by applying $sneg$ to the applicability condition for $+d$ and simplifying
(using the coherence of $\Delta$).

Consider a defeasible theory $D$ consisting of facts $F = \{a, b\}$, and rule $R = \{b \Rightarrow f\}$.
The following is a proof in $\DL(d)$: $+d\, f; +\Delta a; -d\, f$.  Thus the logic is not coherent.

Before $+\Delta a$ is inferred, $+d$ can infer literals not in $F$.
After    $+\Delta a$ is inferred, $-d$ can infer literals not in $F$.
Thus any $+d$-inference made before $+\Delta a$ can be matched with an incoherent inference after $+\Delta a$.
\end{example}

Finally, we present a syntactically-defined subclass of well-behaved logics.
We first need some definitions, to formalise the stratification mentioned in Section~\ref{sect:LAC}.
Given a $J$-closure, we say it \emph{involves} an inference rule $d$
if $d$ is in the smallest reference-closed set containing $J$.
A \emph{closure stratification} is a mapping $m$ from tags to natural numbers such that:
\begin{itemize}
\item
$m(d) \geq m(d')$, whenever an inference rule for $+d$ or $-d$ refers to $+d'$ or $-d'$
\item
$m(d) > m(d')$, whenever an inference rule for $+d$ or $-d$ refers to a closure that involves $d'$
\end{itemize}
A logic is \emph{closure stratified} if it has a closure stratification.

We say a logic is \emph{well-disciplined} if
it is closure stratified,
every inference rule is $P$-disciplined,
it supports the revised Principle of Strong Negation,
and
all pre-defined sets in inference rules are even-handed closures.

\begin{proposition}   \label{prop:cohere}
If a defeasible logic is well-disciplined then it is well-behaved.
\end{proposition}
\skipit{
\begin{proof}
Since the logic is well-disciplined, all inference rules are $P$-disciplined and hence stable,
by Proposition~\ref{prop:stable}.
In addition, the logic supports the revised Principle of Strong Negation,
and
all pre-defined sets in inference rules are even-handed closures.
Thus, it only remains to prove that all closures are coherent.
The proof is by induction on the closure stratification of the tags.

Consider all tags $d$ with $m(d) = 0$.
The corresponding inference rules must use no pre-defined sets.
Consequently, the logic made up only of these inference rules is well-behaved and,
by Theorem~\ref{thm:cohere}, coherent.

Suppose the logic consisting of all inference rules $\pm d'$ wiith $m(d') < n$
is well-behaved and coherent.
Consider all tags $d$ with $m(d) = n$,
Then all closures referred to by $\pm d$ must be coherent.
Consequently, the logic made up only of those inference rules $\pm d$ with $m(d) \leq n$
is well-behaved and,
by Theorem~\ref{thm:cohere}, coherent.

By induction, for every $n$,  the  logic consisting of all inference rules $\pm d'$ wiith $m(d) \leq n$
is well-behaved and coherent.
Since a defeasible logic has only finitely many inference rules,
the well-disciplined logic is well-behaved.
\end{proof}
}

\section{Conclusion}   \label{sect:conc}

We have formulated an umbrella notion of defeasible logics
that presents the basic structure of defeasible logics
and accommodates the features of both $\DL(\partial)$ and $\DL(\pl)$.
However, this umbrella notion includes logics that do not address the traditional concern
of defeasible logics 
(reasoning from defeasible rules in a principled and computationally convenient manner)
and/or do not have desirable properties for defeasible reasoning.
After analysing the situation in detail,
we have proposed a class of defeasible logics that support the traditional aims of such logics,
and retain important properties.
We call this class the well-behaved defeasible logics. 
All logics in this class are coherent.

A key element of the definition of this class is the notion of stability of inference,
a monotonicity property of proofs.
We identified a syntactic subclass of inference rules -- those that are $P$-disciplined -- that ensures stability.
It is currently unclear whether there are interesting stable defeasible logics that are not $P$-disciplined.
Clarity on this point might allow a simplification of the class of defeasible logics,
or might identify defeasible logics substantially different from current logics.

One advantage of this study is
that the defeasible theory $D$ appears only as a parameter to
applicability conditions:  the structure and syntax of $D$ is irrelevant to the results proved.
Consequently,
the results will extend to other forms of defeasible theories,
for example to defeasible theories with more kinds of rules 
(such as modal defeasible logics \cite{modalDL})
and/or additional structures
(such as conflict sets \cite{MN10} or mutexes \cite{mutex})
or, perhaps, defeasible theories allowing arbitrary propositional formulas
in place of literals.

The focus of this paper was traditional defeasible reasoning,
but during the analysis several interesting extensions or variations of defeasible logics cropped up that,
although not addressed in this paper,
are worthy of further investigation:
\begin{itemize}
\item
the use of multiple proofs to provide a model of distributed defeasible reasoning
\item
combining aspects of defeasible and default logics through selective use of
instability in defeasible logics
\item
employing closures involving unstratified inference rules to support reflection in reasoning
\item
treating slots in a proof as resources
(in contrast to linear logic \cite{LinearLogic} in which formulas can be considered a resource)
\end{itemize}

It also remains to investigate the extension of this work to well-founded defeasible logics
and  other variants of defeasible reasoning.

\textbf{Acknowledgements:}
The author has an adjunct position at Griffith University and an honorary position at UNSW.
He thanks Dale Miller for correspondence that helped the author clarify his views.

\bibliographystyle{splncs04}
\bibliography{sdl4}

\end{document}